\documentstyle[prb,aps]{revtex}
\begin{document}
\title{Diffraction of He atoms from Xe monolayer 
adsorbed on the graphite (0001) revisited: The importance of multiple scattering processes}
\author{A. \v{S}iber$^{1,}$\thanks{Corrresponding author. Tel.: +385-1-4698839; fax: +385-1-4698890; E-mail: asiber@ifs.hr} and B. Gumhalter$^{1,2}$ }
\address{$^{1}$Institute of Physics of the University, P.O. Box 304, 10001 Zagreb, Croatia}
\address{$^{2}$The Abdus Salam International Centre for Theoretical Physics, Trieste, Italy }

\date{\today}
\maketitle

\begin{abstract}
We comment and discuss the findings and conclusions of a recent theoretical study of the diffraction of He atoms from a monolayer of Xe atoms adsorbed on the graphite (0001) surface [Khokonov et al., Surf. Sci. 496(2002)L13]. By revisiting the problem we demonstrate that all main conclusions of Khokonov et al. that pertain 
to the studied system are at variance with the available experimental and theoretical evidence and the results of multiple scattering calculations presented in this comment.
\end{abstract}

\vskip 4 cm


\begin{center}
{\bf Published in Surf. Sci. 529, L269 (2003)}
\end{center}
\newpage

In a recent Letter\cite{KKK} and subsequent Erratum\cite{KKKerratum} Khokonov, Kokov and Karamurzov\cite{KKK} (hereafter
to be referred to as KKK) treated the problem of He atom scattering (HAS) from
an ordered monolayer of Xe atoms adsorbed on the (0001) surface of graphite (Gr).
The authors discussed a specific model of phonons in the Xe monolayer
and used it in a scattering
calculation based on the hard-wall model and the eikonal approximation with the aim to reproduce and interpret the experimental He atom diffraction intensities reported by Bracco
et al.\cite{Bracco1}.
However, as we shall show, neither their description of phonons in the Xe
overlayer, nor the results of calculations for the scattering
intensities are consistent with the previous
experimental and theoretical studies of the system Xe/Gr(0001) and the  results of revisited calculations described below.

The measurements of diffraction intensities from He beams incident normal to a $(\sqrt{3}\times\sqrt{3})R 30^{0}$ monolayer lattice of Xe atoms on Gr(0001) carried out by Bracco et al.\cite{Bracco1} indicated large corrugation amplitudes as probed by HAS. These authors have attempted to interpret the measured diffraction spectra by using the hard corrugated wall (HCW) to model surface corrugation amplitudes obtained from pairwise summation of atomic He-Xe potentials\cite{Scoles}, and the eikonal approximation \cite{Garibaldi} to calculate the diffraction intensities pertinent to the thus constructed HCW. However, their calculations showed that a more realistic scattering model was needed to obtain a better agreement between the experimental data and theoretical results.

The studies of phonons in a Xe monolayer on Gr(0001) surface by HAS have been presented in Ref. \onlinecite{Xegraphphon}. These
measurements have demonstrated that the scattered He atoms couple most strongly to a
low energy dispersionless or Einstein-like mode with frequency of $\sim$ 3.2
meV. Both the lattice dynamical calculations \cite{Dero} and
the molecular dynamics simulations \cite{marchese} utilising 
Xe-Xe potentials known from the gas-phase have shown that 
this nearly dispersionless mode corresponds to the vibrations of Xe atoms 
that are polarised {\em perpendicular} (vertical) to the surface plane, i.e. to a so called FT$_{z}$ or S-phonon mode localised in the adlayer 
(for illustration of dispersion of all three Xe adlayer
modes see Fig. 4 of Ref. \onlinecite{marchese}). Low energy modes
of the same S-character have been observed and measured also in Xe monolayers on Cu
\cite{XeCu}, Pt \cite{XePt}, Ag \cite{Sibener} and NaCl \cite{XeNaCl}
surfaces. 
It has also been shown\cite{Xegraphphon,XeCu} that a mode of this character hybridises with the substrate
Rayleigh wave mode only for the two-dimensional phonon wavevectors restricted to a small region around the avoided crossing of the two dispersion curves in the first surface Brillouin
zone (SBZ) where the S-mode polarisation vector is no more strictly localised in and perpendicular to the adlayer.

In their treatment of Xe overlayer phonons, KKK have assumed a rigid
Gr substrate and in this approximation the S-mode remains localised in
the adlayer for all phonon wavevectors in the SBZ. Their calculations predict that in Xe/Gr(0001) the vertically polarised S-mode exhibits dispersion from the value of 0.8 meV at the centre of the first SBZ up to 4.5 meV at the zone edge (see Fig 2. of Ref. \onlinecite{KKK}).
A molecular dynamics simulation of the same system, that was also based on the assumption of a rigid substrate, was presented in Ref. \onlinecite{marchese} and yielded an almost dispersionless S-mode with vertical polarization, in accord with the  experimental evidence\cite{Xegraphphon}. Lattice dynamics 
calculations of similar Xe adlayer systems, i.e. Xe/Cu(100) and Xe/Cu(111), have been presented in Ref. \onlinecite{XeCu}. There, it was found first experimentally, and then confirmed theoretically
that also in these systems the adlayer S-mode is to an excellent approximation completely dispersionless. 
Hence, all the results pertaining to vibrations of Xe adlayers on Gr(0001) and several other substrates are at strong variance with the first finding of KKK who obtain that the S-mode in Xe/Gr(0001) exhibits strong dispersion. This type of acoustic-like dispersion is typical
of the adlayer modes with longitudinal (L) and shear horizontal (SH) polarisation that couple much weaker to the scattered
He atoms\cite{Gumarep} and hence their presence cannot explain why in the experiments of inelastic HAS from Xe/Gr(0001) surface\cite{Xegraphphon} the coupling is strongest
to an almost dispersionless mode with a frequency of 3.2 meV.
Hence, as a consequence of the inappropriate description of Xe adlayer in Ref. \onlinecite{KKK}, the density of states they obtain for the
S-mode displays a Debye-like behaviour (Fig 3.a of Ref. \onlinecite{KKK}),
whereas it should follow the density of states characteristic of a flat,
dispersionless mode. Thus, the mean square
amplitudes of Xe atoms KKK could derive from such an inadequate 
density of states (Fig 3.b of
Ref. \onlinecite{KKK}) would be generally wrong, although their magnitudes in a particular case may be
close to the correct ones due to
the fact that the very different models of vibrations can produce similar mean
square displacements.

Although the above described erroneous premises and lattice dynamics 
description of Xe/Gr(0001) employed by KKK suffice
to disqualify all the main conclusions of their analysis\cite{KKK}, we also show that
their scattering calculation is inadequate and hence cannot provide a realistic
description of the collision system He$\rightarrow$Xe/Gr(0001).

KKK have used\cite{KKK} an eikonal-like semiclassical
expression\cite{Garibaldi} to calculate the scattering amplitudes in He$\rightarrow$Xe/Gr(0001) collisions and in this approach the Xe/Gr(0001) surface was
modelled by a corrugated hard wall without introducing Beeby's correction to partly compensate the neglect of the potential well in front of the surface. However, the applicability of this approach suffers from two fundamental restrictions\cite{HCW} which were already noted and discussed in Ref. \onlinecite{Bracco1}. First, the eikonal approximation is known to neglect the multiple scattering processes and therefore cannot be used in the cases of large surface corrugations and low projectile incident energies in which multiple scattering becomes important. Second, the calculations based on the HCW model can serve only as a rough estimate of the corrugation amplitudes and cannot be used as a tool for quantitative determinations of the
surface structure and the details of interaction potential. 
These restrictions become especially important in the case of strongly corrugated surface profiles as is the case with the present system (see
Fig. \ref{comKKKfg1}). In particular, if the repulsive part of He-target potential is treated
as an infinitely steep wall it is known that it cannot realistically represent the features of
He-target interactions\cite{Vidali}. Also, since in the HCW model the attractive component of the interaction is completely neglected, this
approximation cannot be expected to hold at low incident He atom energies. 
Even for the room temperature He beam energies (the case
considered by KKK), for which $E_i$=63.8 meV, this approximation cannot 
be expected to provide reliable values for the intensities of diffraction
peaks corresponding to He atoms scattered into the states with low
energy in the $z$-direction (perpendicular to the surface). 
The well depth $D$ of the He-Xe/graphite potential is known with high precision from the 
studies reported in Refs. 
\onlinecite{Hutson} and \onlinecite{Aziz} and amounts to
$D=7.4$ meV, and hence the He atoms scattered into the (50) and
(60) channels (c.f. Figs. 1 and 4 of Ref. \onlinecite{KKK} and Fig. 2 of
Ref. \onlinecite{Bracco1}) have components of energy in the $z$-direction of
26 meV and 9.4 meV, respectively, that are comparable to $D$.
To remedy the shortcomings of the HCW model in applications to the
He$\rightarrow$Xe/Gr(0001) system, KKK have assumed\cite{KKK,KKKerratum} the magnitudes of corrugation parameters of the HCW shape function $\zeta({\bf R})$ for modelling the Xe/Gr(0001)  surface that are somewhat different from the best fit parameters suggested by Bracco et al.\cite{Bracco1} and, moreover, are inconsistent with the corrugation of the isopotential surface at 64 meV that can be constructed from the best available He-Xe gas phase binary potentials (c.f Fig. \ref{comKKKfg1}). 
Also, the HCW shape function employed by KKK\cite{KKK,KKKerratum} in their  eikonal approximation calculation of the scattering intensities, despite having the same functional form as the one given by Bracco et al., may appear as a possible source of computational errors because it is manifestly expressed in terms of the rectangular rather than oblique Cartesian coordinates 
introduced in Ref. \onlinecite{Bracco1} (c.f. Fig. 1 of Ref. \onlinecite{KKK} and inset in Fig. 2 of Ref. \onlinecite{Bracco1}).

To quantitatively substantiate the inadequacy of the scattering model on
which the calculations in
Refs. \onlinecite{KKK} and \onlinecite{KKKerratum} had been based, we have performed a coupled channel calculation\cite{Wolken} (CC) of the diffraction intensities
for helium atom scattering from Xe/Gr(0001) surface\cite{SiberPhD} using
the realistic He-target potentials known from the literature. We treat
the target as static and construct the total He-Xe/Gr(0001) potential as a sum 
of the pairwise He-Xe interactions known from the gas phase\cite{Cvetko} and 
the long range interaction of He with the graphite 
substrate\cite{Vidali,Hutson}. This potential  is very
similar to the one used by Hutson and Schwartz\cite{Hutson}, and the very small differences are due to the fact that the He-Xe gas-phase 
potential we use is the one suggested by Cvetko et al. \cite{Cvetko},
whereas Hutson and Schwartz\cite{Hutson} used a slightly different gas-phase
He-Xe potential. 
The obtained total potential gives rise to the equipotential surface at 64 meV which exhibits the peak-to-peak corrugation amplitudes $\chi_1=0.74$ \AA\hspace{1mm}  
and $\chi_2=0.96$ \AA\hspace{1mm}  
in the two high symmetry directions along the surface (see inset in Fig. \ref{comKKKfg1}). A summary of peak-to-peak corrugation amplitudes characteristic of the HCW models of Refs. \onlinecite{KKK} and \onlinecite{KKKerratum}, of Ref. \onlinecite{Bracco1}, and of the ones obtained from the present calculation is given in Table 1. 
Note here that in order to obtain the amplitudes given in the first row of this Table we have corrected   
two errors appearing in the HCW shape function $\zeta_{0}({\bf R})$ quoted in Eq. (7) of Ref. \onlinecite{KKK}, 
viz. the misprinted value of the parameter $\zeta_{10}$ 
taken from Ref. \onlinecite{Bracco1}, which should read 
$\zeta_{10}=0.098$ \AA (c.f. Ref. \onlinecite{KKKerratum}), and the role of $x$ and $y$ coordinates in the 
shape function of Ref. \onlinecite{KKK} which were here taken to have the same 
meaning of {\em oblique} coordinates as in Ref. \onlinecite{Bracco1} (c.f. the discussion at the end of the preceding paragraph).

The CC calculations with a static periodic potential yield  intensities of the $\delta$-function-like diffraction peaks and in order to facilitate comparison with experiments we have broadened the calculated $\delta$-functions by Gaussians of width 
$\sigma_{\theta_f}$ to account for the finite energy spread of the incident beam, 
again given by a Gaussian of width $\sigma_E$. It
is straightforward to show that the above Gaussian width parameters are 
related to each other, and for normal projectile incidence this relation reads:

\begin{equation}
\sigma_{\theta_f} = \frac{1}{2 \cos \theta_f} \frac{G_f}{k_i}
\frac{\sigma_E}{E_i},
\end{equation}
where $G_f$ is the inverse lattice vector associated with a transition 
into a particular diffraction channel, $k_{i}$ is the projectile initial
wave vector ($k_{i} = \sqrt{2mE_{i}}/\hbar$, $m$ is the projectile mass) and
$\theta_{f}$ is the final scattering angle with respect to the $z$-axis
(in our calculations, we have taken $\sigma_{E} / E_{i}=7$ \%). 

In Fig. \ref{comKKKfg2} we present the results of our CC calculations (full thick line in the lower panel) in
comparison with the experimental data of Bracco et al.
\cite{Bracco1} (upper panel). 
As can be seen from Fig. \ref{comKKKfg2}, the agreement with the experimental data is
very satisfactory as regards the relative intensity ratios. The calculated  relative intensities are also in a very good agreement with the results of earlier calculations by Hutson and Schwartz\cite{Hutson}.
To illustrate the importance of the attractive
component of the interaction that was completely ignored in the
calculations of KKK, we also display the results of independent CC
calculations in which either of the two important features associated with the existence of the potential well and modifying the multiple scattering effects is neglected, i.e. (i) the exclusion of closed channels (evanescent waves) from the CC basis, or (ii) the exclusion of attractive component of the total potential from the calculation.  As can be seen from the comparison of the full and dash-dotted line in lower panel of Fig. \ref{comKKKfg2},
the first effect caused by the exclusion of closed channels is large for the (60) diffraction
peak whose intensity turns out to be more than a factor of 2 weaker in this
approximation. However, the second effect of exclusion of the attractive potential 
is much more important since it influences the open diffraction channels
in a more profound way and not only through the closed scattering
channels. This effect is demonstrated by carrying out the CC calculations in which all the Fourier components of the attractive He-Xe/Gr(0001) interaction potential are neglected. The resulting diffraction spectrum shown by the dashed line in lower panel of Fig. \ref{comKKKfg2} is very different from the one obtained with the full potential and does not reproduce the experimental spectrum in any of the important aspects. This is in full accord with Ref. \onlinecite{Hutson} where it has been shown that even small variations of the scattering potential can give rise to very large variations of the diffraction intensities. 
Hence, the recalculated CC-intensities that make use of the otherwise realistic He-Xe potentials with the attractive component removed produce the results which strongly differ from the ones obtained with the complete potential. This clearly demonstrates the inadequacy of the various approximate schemes based on the neglect of the effects of attractive components of the full potential in applications to He atom diffraction in the scattering regime in which the measurements reported in Ref. \onlinecite{Bracco1} were carried out.    
In other words, the relatively moderate differences among peak-to-peak corugation amplitudes displayed in Table 1 may be deceptive as regards the behaviour of calculated diffraction intensities because they may give rise to different effects and results in different scattering models.

The overall intensities of diffraction peaks obtained by the CC
calculation with the full potential are about a factor of 10 larger than those obtained
experimentally. This suggests that about 90\% of He atoms in the
experiment of Bracco et al.\cite{Bracco1} were scattered inelastically. Indeed, the experimental intensities when summed over all the measured 
diffraction  peaks (c.f. Table 1 of Ref. \onlinecite{Bracco1}) yield only 11.2
\% of the incident intensity, which means that nearly 88.8 \% of the scattered He atoms  end up in inelastic channels, in excellent agreement with the
factor of 10 that can be deduced from comparison of the results of our
calculations and experiments. Note that this also implies a multiphonon
scattering regime since the mean number of phonons $\overline{n}$ excited
in a scattering event can be estimated from $\overline{n}_{exp}=-\ln
(I/I_0)=-\ln(0.11)=2.21$, where $I$ and $I_{0}$ are the experimental values of the total intensity scattered into the elastic channels and the incident beam intensity, respectively, and $I/I_{0}$ is the Debye-Waller factor (c.f. Ref. \onlinecite{Gumarep}). It should be
pointed out that this finding is again at variance with the conclusion of Ref. \onlinecite{KKK}, that "the multiphonon processes can be neglected". 
Namely, KKK have arrived at such a conclusion by uncritically employing Weare's criterion\cite{Weare} for $\overline{n}$ in the regime in which this criterion is inapplicable (c.f. Figs. 1 and 2 in Ref. \onlinecite{Sibrev}) and, moreover, made a numerical error in their estimate. To clarify this issue we have also calculated
the theoretical value $\overline{n}_{th}$ for the present system using the EBA multiphonon scattering formalism
described in detail in Ref. \onlinecite{Gumarep}. This yields
$\overline{n}_{th}=2.6$, again in a very good agreement with the above discussed value of $\overline{n}_{exp}$ 
that can be extracted from experiments. 

In summary, by making use of the available He-Xe atomic pair potentials
(Ref. \onlinecite{Cvetko}) and the results of experimental
investigations of the Xe/Gr(0001) surface by HAS
(Refs. \onlinecite{Bracco1} and \onlinecite{Xegraphphon}), the coupled
channel method\cite{Wolken} for calculation of the diffraction intensities in thermal
energy atom scattering from corrugated surfaces, and of the multiphonon
scattering formalism\cite{XeCu,Gumarep} in combination with the lattice and molecular
dynamics analyses\cite{Dero,marchese} to calculate the Debye-Waller attenuation of diffraction intensities, we have
demonstrated that the results and conclusions reached by the authors of
Ref. \onlinecite{KKK} and pertaining to: (i) the dispersion of vertically
polarised modes in  the Xe
monolayer, (ii) the properties and modelling of the He-Xe/Gr(0001) interaction
with the relatively large potential well and surface corrugation profile, and (iii) the role of
multiple scattering processes in thermal energy HAS from Xe/Gr(0001) surface, are either inadequate or in a complete disagreement both with the existing experimental
evidence and with the theoretical assessments of  the
He$\rightarrow$Xe/Gr(0001) collision system.
The arguments which we supply to refute the conclusions reached in
Ref. \onlinecite{KKK} show that: (i) the vertically polarised vibrational mode in Xe
monolayer on Gr(0001) surface is
nearly dispersionless, i.e. it is a flat S-mode which due to its low excitation frequency and perpendicular to the surface polarisation strongly couples to the scattered He atoms, (ii) the potential well depth and the surface corrugation
amplitudes characteristic of the He-Xe/Gr(0001) interaction under the studied scattering conditions are large, which necessitates a self-consistent treatment of the diffraction intensities within a multiple scattering formalism
(and therefore any agreement between the results of the diffraction intensity calculations of Ref. \onlinecite{KKK} and experimental data can
be only accidental), and (iii) the multiphonon
processes arising from the strong He atom-S-mode coupling play an important role in the Debye-Waller factor-induced attenuation of experimental diffraction
peak intensities reported in Ref. \onlinecite{Bracco1} with which the authors of
Refs. \onlinecite{KKK} and \onlinecite{KKKerratum} have compared their calculated intensities.


\begin{figure}
\caption{Contour plot of the He-Xe/Gr(0001) interaction potential 
$V({\bf R},z)$ (in meV) along two high surface symmetry lines (a) and (b) shown 
in the inset. The energies of the isopotential lines are given in units of meV. The thick contour denotes the isopotential surface for 
which $V({\bf R},z)=E_{i}=64$ meV.}
\label{comKKKfg1}
\end{figure}

\begin{figure}
\caption{Top panel: Experimental diffraction intensities for He atom scattering from Xe/Gr(0001) surface shown as a function of the final 
scattering angle (after Ref. \protect\onlinecite{Bracco1}). Scattering parameters given in the inset. Bottom panel: Full line: Results of converged CC calculations using the total He-Xe/Gr(0001) potential as described in
the text. Dash-dotted line: Results of CC calculations
excluding closed diffraction channels. Dotted line: Results of CC calculations using the He-Xe/Gr(0001) potential without the attractive components.}
\label{comKKKfg2}
\end{figure}

\begin{table}
\begin{tabular}{|c|c|c|}
Reference & $\chi_{1}$ [ \AA ] & $\chi_{2}$ [ \AA ] \\
\hline
KKK\protect\cite{KKK,KKKerratum} & 0.72 & 0.88 \\
\hline
Bracco et al.\protect\cite{Bracco1} & 0.64 & 0.97 \\
\hline
This work & 0.74 & 0.96 
\end{tabular}
\caption{First and second row: peak-to-peak HCW corrugation amplitudes used to compute He beam diffraction intensities in Refs. \protect\onlinecite{KKK} and \protect\onlinecite{KKKerratum}, and in Ref. \protect\onlinecite{Bracco1}, respectively. Third row: the corresponding amplitudes derived in the present work from the classical turning points in the He-Xe/Gr(0001) interaction potential $V({\bf R},z)$ for He atom incident energy of 64 meV (see Fig. \protect\ref{comKKKfg1}).}
\end{table}

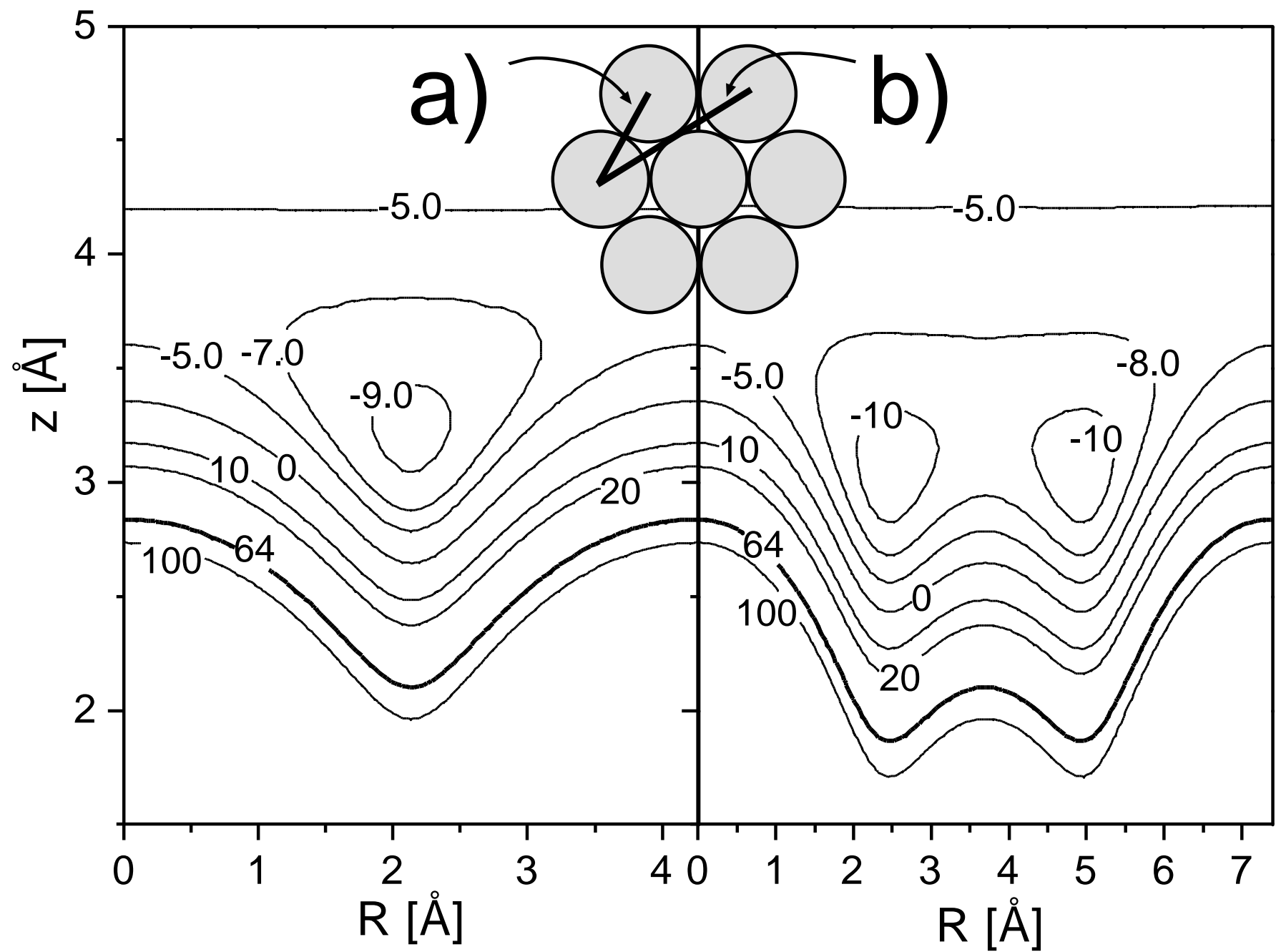
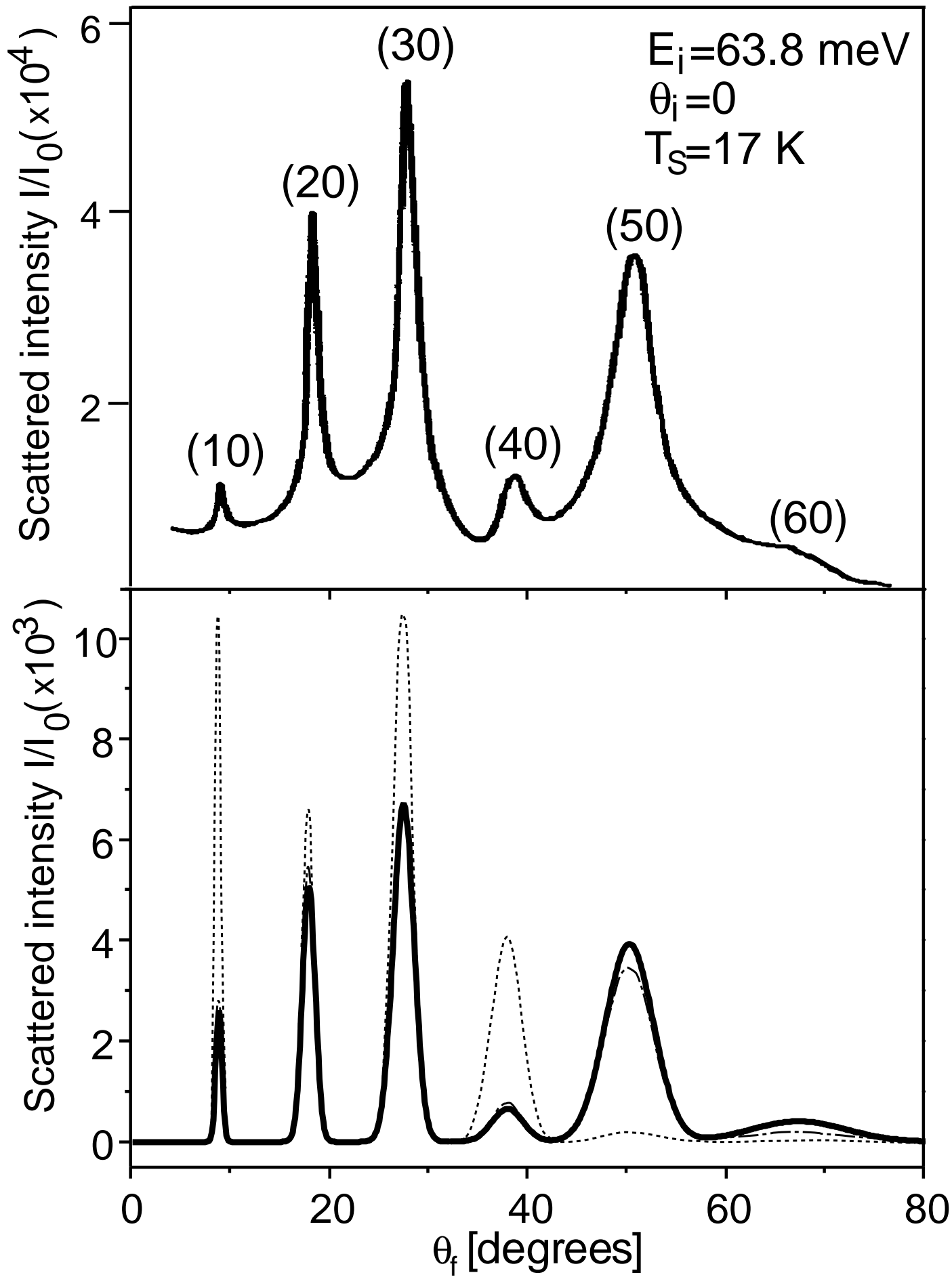
\end{document}